\documentclass[amssymb,aps,twocolumn,superscriptaddress,floatfix,longbibliography, nobalance]{revtex4-2}
\usepackage{amsmath}
\usepackage{mathrsfs} % for \mathscr
\usepackage{graphicx} % for figures
\usepackage{comment} % allows block comments
\usepackage{xcolor} % allow colour change in the text
\usepackage[normalem]{ulem} % allows strikeout text, e.g. \sout{text}
\usepackage{braket}
\usepackage{subcaption}
\usepackage{tikz}
\usepackage{tikz-3dplot}

\begin{document}

\title{No violation on a generalisation of Leggett-Garg inequality and Bell-CHSH inequality with extended probability}

\author{Sirawit Kajonsombat}
\email[email: ]{sirawit.kajonsombat@gmail.com \space (corresponding author)}
\affiliation{The Institute for Fundamental Study (IF), Naresuan University, 99 Moo 9, Tah Poe, Mueang
Phitsanulok, Phitsanulok, 65000, Thailand}
\author{Pongwit Srisangyingcharoen}
\email[email: ]{pongwits@nu.ac.th}
\affiliation{The Institute for Fundamental Study (IF), Naresuan University, 99 Moo 9, Tah Poe, Mueang
Phitsanulok, Phitsanulok, 65000, Thailand}

\begin{abstract}
    A generalisation of the Leggett-Garg inequality (LGI) and the Bell-CHSH inequality is proposed in this work, by replacing a classical probability notion with an extended probability notion. The extended probability serves as an underlying layer of reality to the classical probability. Within the consistent history framework, the underlying layer is so-called a non-settleable history. Hence, the bounds of the LGIs and Bell-CHSH inequalities are extended. The resulting generalised inequalities are satisfied without violation for arbitrary measurement settings. Furthermore, generalised Macrorealism and generalised Local realism are introduced and analysed within the extended probability framework without relying on an explicit measurement. This opens up further possibilities on one of the great pursuits in physics, understanding on how a quantum system can possess classical properties.
\end{abstract}

\maketitle

\section{introduction}
\indent Understanding on how to reconcile quantum reality with classical reality, a common playground where both exist, is one of the great pursuits in physics. One of the most successful explanations to this problem is decoherence \cite{zurek1991decoherence}. However, there remains separation between quantum and classical regimes. Several approaches have been developed to describe how a classical description can emerge from a quantum system, including the consistent history (CH) framework \cite{dowker1992quantum,landsman2006between}. In this framework, a world-view of a closed system comprises both quantum and classical property \cite{gell1996quantum, gell1993classical, halliwell1994review, griffiths2003consistent, hartle2016decoherent, griffiths2013consistent}. The CH framework allows one to have a quantum system such that it is compatible with classical property without any separation, with a starting belief that quantum property is the fundamental property of nature. The framework considered in this work will rely on this world-view thoroughly.
\\
\indent The Macrorealism (MR) and Local realism (LR) are the classicality of interest. A system that possesses either MR or LR is a definitive system in which there is no notion of superposition of observed properties and the measurement independence is preserved. This is a typical behaviour of a classical system we used to. Additionally, while LR constrains spatial correlations between separated subsystems, MR constrains temporal correlations of a single system over time.
\\
\indent It is known that a quantum system violates both the Leggett-Garg inequality (LGI) \cite{leggett1985quantum} and the Bell-CHSH inequality \cite{bell1964einstein,clauser1969proposed}, therefore the system cannot possess MR and LR under their conventional definition. In other words, these inequalities require the existence of a non-negative joint probability space which is not generally possible for quantum systems \cite{hess2005bell, maccone2013simple}. In this work, we will relax this requirement by exploiting an extended probability \cite{hartle2004linear, goldstein1995linearly} in the LGI and Bell-CHSH inequality, similarly to what have done in \cite{halliwell2013negative, cereceda2000local, rothman2001hidden, halliwell2016leggett} while discussing on the consequences of no violation. These honourable works \cite{morris2022witnessing, emary2017ambiguous} are comparable to our model, however, we are not interested in retaining the classical probability space but rather exploiting the consequences of allowing an extended probability.
\\
\indent The structure of this paper will be as follows: A brief review of necessary concepts are given in section II, including an essential concept of extended probability in the history formalism by J.B. Hartle \cite{hartle2008quantum}. No-violation of a generalised LGI and the reconsideration on the generalised Macrorealism are discussed in section III. Moreover, in section IV, generalised Bell-CHSH inequality and generalised Local realism will be examined in a similar way. An additional remark is added in section V. We close the presented work with a conclusion in section VI.

\section{Review of necessary concepts}
\subsection{Leggett-Garg inequality}
\indent Leggett-Garg inequality (LGI) is a set of inequality that provides a test of Macrorealism \cite{leggett1985quantum, fritz2010quantum, emary2014leggett, halliwell2019fine, vitagliano2023leggett}. Consider a four-time measurement Leggett-Garg inequality (LGI4) which is defined as
\begin{equation}
    K_4 \equiv \left| C_{s_1 s_2} + C_{s_2 s_3} + C_{s_3 s_4} - C_{s_1 s_4} \right| \le 2.
\end{equation}
where each temporal correlation is defined by
\begin{equation}
    C_{s_is_j} = \sum_{s_is_j} s_i s_j p(s_i, s_j).
\end{equation}
and $s_i = \pm1$ are possible results at each time $t_i$. A two-time classical probability $p(s_i,s_j)$ has the form
\begin{equation}
    p(s_i,s_j) = \text{Tr}(P_{t_j} P_{t_i} \rho P_{t_i})
\end{equation}
that satisfies both $p \ge 0$ and $\sum p =1$ where $P_{t_i}$ is the projection operator at a given time $t_i$. 
\\
\indent There are three essential assumptions for a system to possess MR and necessarily satisfy LGI:
\begin{itemize}
    \item Macrorealism per se (MRps) --- An observed state at any given time must be in one of the available states.
    \item Non-invasive measurement (NIM) --- It is possible in principle to perform a measurement in such a way that neither the state at a given time nor its subsequent state is influenced by the measurement.
    \item Induction --- A future measurement, whether measured or not, should have no effect on the current state. This assumption is always assumed to hold.
\end{itemize}
However, according to \cite{halliwell2019necessary}, the set of inequalities alone is only a necessary condition for MR. Particularly the two-time marginal probability in Eq.(3) cannot be obtained from a non-negative joint probability due to the absence of Fine's theorem \cite{fine1982hidden, fine1982joint}. Hence, a No-signalling in time (NSIT) condition \cite{kofler2012condition,clemente2015necessary,clemente2016no}, which is also a necessary condition that characterises NIM property, reads
\begin{equation}
    \sum_{s_j} p(s_i,s_j) = p(s_i).
\end{equation}
Thus, if a non-negative joint probability satisfies a set of NSIT conditions, then the NIM also holds. However, a classical probability in a quantum system does not generally satisfy the NSIT conditions mathematically \cite{halliwell2017comparing,halliwell2019necessary}. Moreover, violation of the inequality implies that both MRps and NIM are violated simultaneously. As is well known, quantum system can violate such assumptions exactly.
\\
\indent As mentioned by \cite{halliwell2019necessary}, the Macrorealism we pursuit here is also the traditional notion of Macrorealism. The observed result should be definite and the measurement should be non-invasive. For other kinds of the Macrorealism and their related theorems, see Ref. \cite{maroney2014quantum}. Additionally, the induction assumption or sometimes known as the arrow of time, either name is a nomenclature problem, is always taken to hold. It states that the future measurement should have no effect on the present one whatsoever. Since extended probability has no explicit relation to deviate the temporal ordering of events, we take the induction assumption for granted as well. Thus, the analysis of this work is concentrated on MRps and NIM assumptions.

\subsection{Bell-Clauser–Horne–Shimony–Holt inequality}
\indent Bell-CHSH inequalities is set of sufficient conditions that is utilised to test Local realism (LR) \cite{bell1964einstein,clauser1969proposed, hohenberg2010colloquium, guimaraes2024introduction}. The Bell-CHSH inequality reads
\begin{equation}
    B_4 \equiv \left| C_{ab} + C_{ab'} + C_{a'b} - C_{a'b'} \right| \le 2
\end{equation}
where $(a,a')$ and $(b,b')$ are the measurement angles of two subsystems, Alice ($A$) and Bob ($B$) respectively. Each spatial correlation is defined as
\begin{equation}
    C_{s_a s_b} = \sum_{s_a s_b} s_a s_b \; p(s_a, s_b).
\end{equation}
where $s_a = \pm1$ and $s_b=\pm1$. A classical probability $p(s_a,s_b)$ is defined as
\begin{equation}
    p(s_a,s_b) = \text{Tr} \left[ (P_{s_a} \otimes P_{s_b}) \rho \right]
\end{equation}
that satisfies $p \ge 0$ and $\sum p = 1$. A system that satisfies Bell-CHSH inequality and admits Local realism that contains two assumptions:
\begin{itemize}
	\item Determinism --- A system possesses a definite value of a measurement result upon whether the measurement is performed or not.
	\item Locality --- A measurement of one subsystem should not affect the other spatially separated subsystem.
\end{itemize}
 Violation of Bell-CHSH inequality is interpreted differently amongst physicists. The mainstream perspective is that the violation implies the rejection of LR world-view entirely. However, in an aspect of probabilists, such a violation is expected since there should be no non-negative classical joint probability in general that satisfies Bell-CHSH inequality \cite{hess2005bell, maccone2013simple, cereceda2000local, rothman2001hidden}. This view is supported by the well-known Fine's theorem \cite{fine1982joint,fine1982hidden,halliwell2019fine}. From the perspective of the consistent history framework \cite{griffiths2003consistent, hohenberg2010colloquium, halliwell1994review}, the hidden variable assumption is even inconsistent since the beginning because it violates the single-framework rule \cite{griffiths2020nonlocality}. 
 \\
 \indent Note that No-signalling (NS) theorem is always presumed to hold to respect the causality. This contrasts to the NSIT in the LGI context. One of the key differences between the two concepts is that the NSIT is falsifiable whereas the NS is not.

\subsection{Extended probability}
\indent Extended probability or quasiprobability has several distinct definitions. The prime example is the one of Wigner distribution. However, the extended probability by Goldstein and Page \cite{hartle2004linear,halliwell2013negative} is of interest in this work, specifically in the context of consistent history. It is defined as
\begin{equation}
    q (\alpha) = \text{ReTr}\left( C_\alpha \rho \right)
\end{equation}
where 
\begin{equation}
    C_\alpha \equiv P^n_{\alpha_n}(t_n)...P^1_{\alpha_1}(t_1) 
\end{equation}
is the history chain operator. Extended probability possesses a time-neutral property, whereas the classical probability,
\begin{equation}
    p(\alpha) = \text{Tr} \left( C_\alpha ^\dagger \rho C_\alpha \right),
\end{equation}
is time-asymmetric by definition. Moreover, the relation between extended probability and classical probability reads
\begin{equation}
    q(\alpha) = p(\alpha) + \sum_{\alpha' \ne \alpha}\text{Re}D(\alpha,\alpha')
\end{equation}
where the decoherence functional, or so-called interference term \cite{majidy2021detecting}, is
\begin{equation}
    D(\alpha,\alpha') = \text{Tr} \left( C_{\alpha} \rho C^{\dagger}_{\alpha'} \right).
\end{equation}
It plays an important role in recovering a classical probability through decoherence when all the decoherence functional vanish,
\begin{equation}
    q(\alpha) = p(\alpha).
\end{equation}
This is known as medium decoherence condition though this is not the only way to remove the interference terms. The other conditions will be given in Sec.II.D. 
\\
\indent Extended probability thus may be useful as an intermediate tool to calculate classical probability; further details can be found in \cite{scully1994feynman, feynman2012negative, hartle2008quantum}. Additionally, in the context of extended probability, MRps could be tested directly by the condition $q \ge 0$ if one believes that Fine's theorem is all there is in quantum system \cite{halliwell2013negative,majidy2021detecting}. However, This is not the case of extended probability we want to emphasise here.

\subsection{Consistent history}
In a Consistent history (CH) framework, a history is represented by a sequence of projection operators at different times, that is a history chain operator Eq.(9). Projection operators are mutually exclusive and exhaustive within each single-frame
\begin{equation}
    P_\alpha P_\beta = \delta_{\alpha \beta} P_\alpha \quad, \quad \sum_{\alpha} P_\alpha = \mathbb{I}.
\end{equation}
No non-commuting observables are expanded simultaneously at each time. This is known as a \textit{single-framework rule}. A chain of histories $C_\alpha$ is thus consequently exhaustive ($\sum_\alpha C_\alpha = \mathbb{I})$, however, it is not necessary exclusive.
\\
\indent The history framework is describable by the settleable and non-settleable histories which are observable or not depending on whether a classical probability can be assigned \cite{hartle2008quantum}. A settleable history corresponds to a settleable-bet interpretation of classical probability where the outcomes are observable. On the other hand, a non-settleable history, which is not observable, corresponds to a non-settleable-bet interpretation, is of interest to us. Unlike in the standard consistent history \cite{griffiths2003consistent} where no finest-grained history exists, the finest-grained history becomes possible by introducing an extended probability. See more details of this finest-grained history analogous to other frameworks in \cite{hartle2008quantum, hartle2016decoherent}. Extended probability also satisfies the NSIT condition \cite{halliwell2016leggett} mathematically. Therefore, we currently have a two-layer picture in our quantum mechanics model---the more fundamental layer is the non-settleable history while the observable layer is the settleable history.
\\
\indent One of the interesting features of combining the history formalism with extended probability is the analogous picture to the standard quantum mechanics where the Schr\"odinger equation is a main character. There, we have a wavefunction evolving according to the Schr\"odinger equation unitarily. However, as in the Copenhagen interpretation, it collapses into an eigenstate upon measurement, breaking the time-neutral property. In the history picture with extended probability, a non-settleable history set serves as our information container with the extended probability being time-neutral, where the classical probability lacks. In this picture, coarse-graining perform the same tasks as the collapse of wavefunction. It converts a non-settleable history into a settleable one. 
\\
\indent This raises the important question ``Is coarse-graining of histories as unphysical as the collapse of wavefunction?''. We believe that is not the case. Coarse-graining does not make the rest of the possible settleable histories disappear physically and thus the classical probability. Yet it makes what is not observable becomes observable. We are not forced to accept classical probability like a collapse interpretation, we accept the settleable basis by choice. Moreover, according to \cite{griffiths2003consistent, griffiths2013consistent}, there is no collapse interpretation required within the consistent history framework. We measure, or observe, a realisable result, not making one. The remaining possible outcomes neither disappear nor get destroyed.
\\
\indent As mentioned previously, obtaining a settleable history requires the elimination of the interference between histories. There are four decoherence functional conditions that achieve this \cite{halliwell2009partial}. They are listed in order --- medium decoherence, partial decoherence, weak decoherence, and linear positivity:
\begin{align}
    D(\alpha,\alpha') &= 0 \;\;\;\;\; ; \forall \alpha \ne \alpha'
    \\
    \sum_{\alpha'; \alpha' \ne \alpha} D(\alpha,\alpha') &= 0
    \\
    \text{Re} D(\alpha,\alpha') &= 0 \;\;\;\;\; ; \forall \alpha \ne \alpha'
    \\
    \sum_{\alpha'} \text{Re} D(\alpha,\alpha') &\ge 0
\end{align}
together with $\sum_\alpha q(\alpha) = 1$. These decoherence conditions work well for a single system (either open or closed). Composite system is a different story, as pointed out by Di\'osi \cite{diosi2004anomalies}. The Di\'osi tests determine whether the decoherence conditions are applied for composite systems. Up to date, the only condition which passes the tests is the medium decoherence.
\\
\indent In summary, from the suggested causes of violation of both LGI and Bell-CHSH, together with the fact that projection operators of non-commuting observables could lead to extended probability while still respecting the single-framework rule of the consistent history formalism, a combination of ideas above offers a way to recover classical perspective within quantum systems. Along with ensuring no violation of such generalised inequalities, this requires a reconsideration of the assumptions underlying MR and LR once the extended probability is involved. 

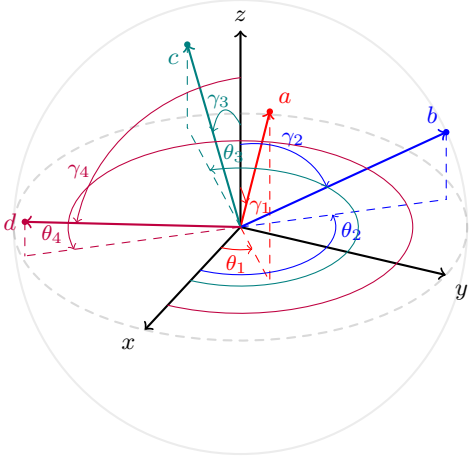
\begin{figure}
    \centering
% View angle: 60 degrees down from z-axis, rotated 115 degrees
    \tdplotsetmaincoords{60}{115}

    \begin{tikzpicture}[tdplot_main_coords, scale=1.2, line cap=round, line join=round]

    % Style for labels: NO background (removed fill=white)
    % inner sep creates a small invisible buffer so the text doesn't touch the line
    \tikzset{lbl/.style={inner sep=2pt, font=\footnotesize}}

    % Sphere radius
    \def\rho{2.5}

    % Define variables for all 4 quadrants (theta in xy-plane, gamma from z-axis)
    \def\thetaA{40}  \def\gammaA{30}  \def\rA{0.5}  % a (Red) - Quadrant 1
    \def\thetaB{130} \def\gammaB{70}  \def\rB{1.05}  % b (Blue) - Quadrant 2
    \def\thetaC{220} \def\gammaC{65}  \def\rC{1.3}  % c (Teal) - Quadrant 3
    \def\thetaD{310} \def\gammaD{80}  \def\rD{1.9}  % d (Purple) - Quadrant 4

    % Draw Negative Axes (dashed) to clearly show quadrants
    %\draw[dashed, gray] (-1.5,0,0) -- (0,0,0);
    %\draw[dashed, gray] (0,-1.5,0) -- (0,0,0);

    % Draw Positive Axes (Solid)
    \draw[thick, ->] (0,0,0) -- (2.5,0,0) node[anchor=north east]{$x$};
    \draw[thick, ->] (0,0,0) -- (0,2.5,0) node[anchor=north west]{$y$};
    \draw[thick, ->] (0,0,0) -- (0,0,2.5) node[anchor=south]{$z$};

    % Draw the sphere outline and equator
    \begin{scope}[tdplot_screen_coords]
        \draw[gray, thick, opacity=0.15] (0,0,0) circle (\rho);
    \end{scope}
    \tdplotdrawarc[gray, thick, opacity=0.3, dashed]{(0,0,0)}{\rho}{0}{360}{}{}

    % ==========================================
    % Measurement a (RED) - Quadrant 1
    % ==========================================
    \tdplotsetcoord{A}{\rho}{\gammaA}{\thetaA}
    \draw[thick, red, ->] (0,0,0) -- (A);
    \draw[dashed, red, thin] (0,0,0) -- (Axy);
    \draw[dashed, red, thin] (A) -- (Axy);
    \fill[red] (A) circle (1pt) node[anchor=south west, text=red] {$a$};
    
    % Arcs
    \tdplotdrawarc[->, red]{(0,0,0)}{\rA}{0}{\thetaA}{lbl, anchor=north, text=red}{$\theta_1$}
    \tdplotsetthetaplanecoords{\thetaA}
    \tdplotdrawarc[->, red, tdplot_rotated_coords]{(0,0,0)}{\rA}{0}{\gammaA}{lbl, anchor=north west, text=red}{$\gamma_1$}

    % ==========================================
    % Measurement b (BLUE) - Quadrant 2
    % ==========================================
    \tdplotsetcoord{B}{\rho}{\gammaB}{\thetaB}
    \draw[thick, blue, ->] (0,0,0) -- (B);
    \draw[dashed, blue, thin] (0,0,0) -- (Bxy);
    \draw[dashed, blue, thin] (B) -- (Bxy);
    \fill[blue] (B) circle (1pt) node[anchor=south east, text=blue] {$b$};
    
    % Real arcs
    \tdplotdrawarc[->, blue]{(0,0,0)}{\rB}{0}{\thetaB}{}{}
    % Dummy arcs to position text near the end of the sweep, pushing outward
    \tdplotdrawarc[draw=none]{(0,0,0)}{\rB}{0}{220}{lbl, anchor= west, text=blue}{$\theta_2$}
    
    \tdplotsetthetaplanecoords{\thetaB}
    \tdplotdrawarc[->, blue, tdplot_rotated_coords]{(0,0,0)}{\rB}{0}{\gammaB}{}{}
    \tdplotdrawarc[draw=none, tdplot_rotated_coords]{(0,0,0)}{\rB}{0}{70}{lbl, anchor=south, text=blue}{$\gamma_2$}

    % ==========================================
    % Measurement c (TEAL) - Quadrant 3
    % ==========================================
    \tdplotsetcoord{C}{\rho}{\gammaC}{\thetaC}
    \draw[thick, teal, ->] (0,0,0) -- (C);
    \draw[dashed, teal, thin] (0,0,0) -- (Cxy);
    \draw[dashed, teal, thin] (C) -- (Cxy);
    \fill[teal] (C) circle (1pt) node[anchor=north east, text=teal] {$c$};
    
    % Real arcs
    \tdplotdrawarc[->, teal ]{(0,0,0)}{\rC}{0}{\thetaC}{}{}
    \tdplotdrawarc[draw=none]{(0,0,0)}{\rC}{0}{400}{lbl, anchor=south east, text=teal}{$\theta_3$}
    
    \tdplotsetthetaplanecoords{\thetaC}
    \tdplotdrawarc[->, teal, tdplot_rotated_coords]{(0,0,0)}{\rC}{0}{\gammaC}{}{}
    \tdplotdrawarc[draw=none, tdplot_rotated_coords]{(0,0,0)}{\rC}{0}{90}{lbl, anchor=south, text=teal}{$\gamma_3$}

    % ==========================================
    % Measurement d (PURPLE) - Quadrant 4
    % ==========================================
    \tdplotsetcoord{D}{\rho}{\gammaD}{\thetaD}
    \draw[thick, purple, ->] (0,0,0) -- (D);
    \draw[dashed, purple, thin] (0,0,0) -- (Dxy);
    \draw[dashed, purple, thin] (D) -- (Dxy);
    \fill[purple] (D) circle (1pt) node[anchor=east, text=purple] {$d$};
    
    % Real arcs
    \tdplotdrawarc[->, purple]{(0,0,0)}{\rD}{0}{\thetaD}{}{}
    \tdplotdrawarc[draw=none]{(0,0,0)}{\rD}{0}{580}{lbl, anchor=north east, text=purple}{$\theta_4$}
    
    \tdplotsetthetaplanecoords{\thetaD}
    \tdplotdrawarc[->, purple, tdplot_rotated_coords]{(0,0,0)}{\rD}{0}{\gammaD}{}{}
    \tdplotdrawarc[draw=none, tdplot_rotated_coords]{(0,0,0)}{\rD}{0}{120}{lbl, anchor=east, text=purple}{$\gamma_4$}

    \end{tikzpicture}
    
    \caption{illustrates the simulation on the variable angles $\{ \theta_i,\gamma_j\}$ of the measurement angles $\left(a,b,c,d\right)$ in a three dimensional configuration.}
    \label{fig:B4}
\end{figure}

\begin{figure}[htbp]
    \centering
    % First subfigure
    \begin{subfigure}[b]{0.4\textwidth}
        \centering
        \includegraphics[width=\textwidth]{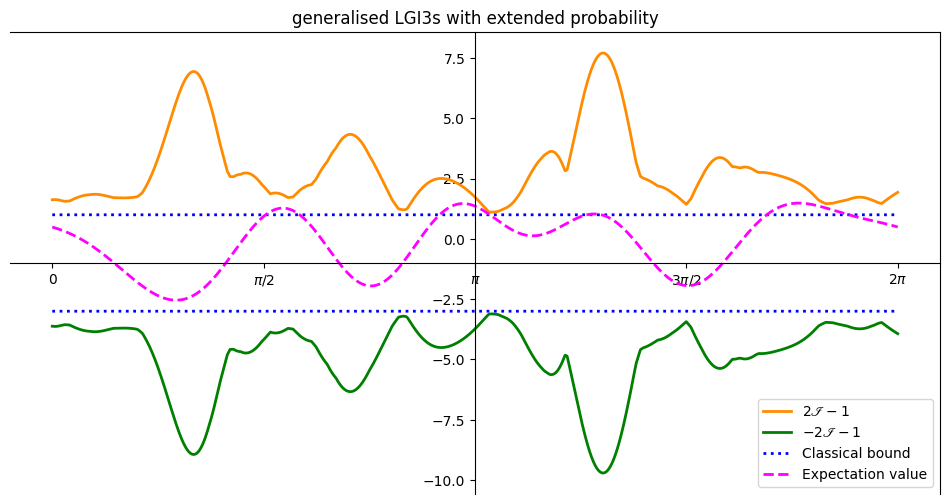}
        \caption{}
        \label{fig:LGI3s}
    \end{subfigure}
    \begin{subfigure}[b]{0.4\textwidth}
        \centering
        \includegraphics[width=\textwidth]{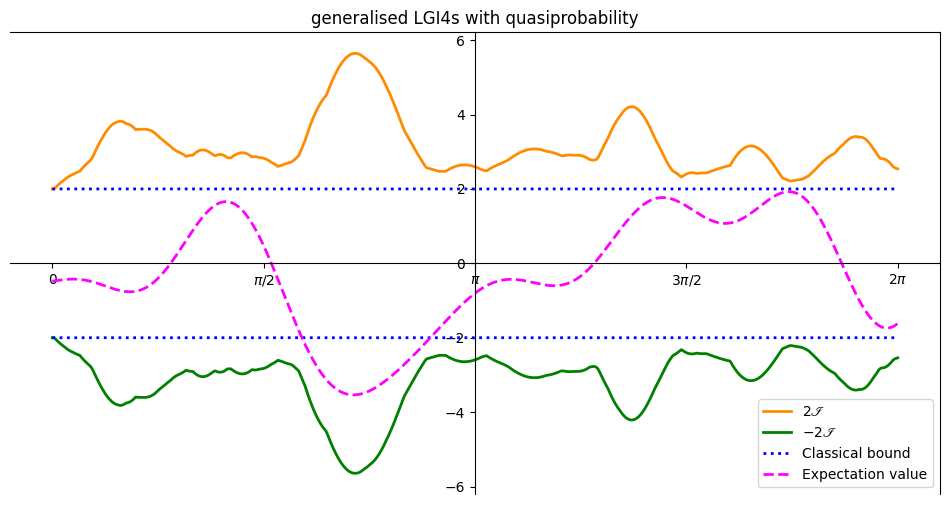}
        \caption{}
        \label{fig:LGI4s}
    \end{subfigure}
    \caption{illustrates a generalised bound for a representative measurement setting in (a) gLGI3 and (b) gLGI4 with extended probability on the spin measurements of a single electron.}
    \label{fig:LGIs_comparison}
\end{figure}

\section{No violation in a generalised Leggett-Garg inequality with extended probability}
\indent We generalised the LGI by replacing a classical probability with the extended probability. The temporal correlation of Eq.(2) is modified to be
\begin{equation}
    \tilde{C}_{s_is_j} = \sum_{s_is_j} s_i s_j \; q(s_i,s_j)
\end{equation}
where the extended probability takes the form
\begin{equation}
    q(s_i,s_j) = \text{ReTr} \left( P_{t_j}^{s_i}P_{t_i}^{s_j} \rho  \right).
\end{equation}
Nevertheless, this modified temporal correlation is numerically equivalent to Eq.(2), as shown in Ref.\cite{halliwell2016leggett}. Therefore, we modify Eq.(1) by replacing temporal correlations. The generalised inequality reads
\begin{equation}
    \tilde{K_4} = |\tilde{C}_{s_1 s_2} + \tilde{C}_{s_2 s_3} + \tilde{C}_{s_3 s_4} - \tilde{C}_{s_1 s_4}|.
\end{equation}
Applying the triangle inequality yields
\begin{align}
    |\tilde{C}_{s_1 s_2}& +  \tilde{C}_{s_2 s_3} + \tilde{C}_{s_3 s_4} -  \tilde{C}_{s_1 s_4}|  \notag \\
    &\le \sum_{s_1 s_2 s_3 s_4}|s_1 s_2 + s_2 s_3 + s_3 s_4 - s_1 s_4| 
    \\ & \quad \quad \quad \quad \cdot |q(s_1,s_2,s_3,s_4)|. \notag
\end{align}
Since the combination of dichotomic operators in Eq.(22) can be either $\pm 2$, the generalised four-time Leggett-Garg inequality (gLGI4) is
\begin{equation}
    |\tilde{C}_{s_1 s_2} + \tilde{C}_{s_2 s_3} + \tilde{C}_{s_3 s_4} - \tilde{C}_{s_1 s_4}|
    \le 2 \mathscr{I}_4
\end{equation}
where we defined the \textit{extended unity}
\begin{equation}
    \mathscr{I}_4 \equiv \sum_{s_1 s_2 s_3 s_4} \left| q(s_1,s_2,s_3,s_4) \right|
\end{equation}
by directly relaxing the condition of a joint probability being non-negative. Note that if a decoherence condition is applied, a usual classical probability boundary  ($\mathscr{I} = 1$) is recovered. This extended unity, which is always $\mathscr{I} \ge 1$, is the main result. Further analysis will be concentrated and demonstrated on this quantity. Note that gLGI3s are retrieved when one of these temporal correlation is either a perfect or anti-perfect correlation. 
\\
\indent A simple model will be exploited to analyse the result, for example, the gLGI4 of a single electron being measured at four times $\{t_1,t_2,t_3,t_4\}$ is being considered here. The projection operator at each time is
\begin{equation}
    P_{t_i}^{s_i} = \dfrac{1}{2} \left[ 1 + s_i (e_i \cdot \sigma) \right]
\end{equation}
where measurement angles $e_1 = a$, $ e_2 = b$, $ e_3 = c$, $ e_4 =d$ are set to be arbitrary in three dimensional space with their variable angles $\{\theta_i,\gamma_i\}$ corresponding to polar angle and azimuthal angle, as shown in Fig.1. The outcomes of the spin operator $\sigma$ is $s_i = \pm1$ as usual. Therefore, the extended probability then reads
\begin{widetext}
    \begin{align}
        q(s_1,s_2,s_3,s_4) &= \mathrm{ReTr} \left(  P_4 P_3 P_2 P_1 \rho \right) \nonumber \\ \nonumber
        &= \frac{1}{16} \{ 1 + s_1 \langle a \cdot \sigma \rangle + s_2 \langle b \cdot \sigma \rangle + s_3 \langle c \cdot \sigma \rangle + s_4 \langle d \cdot \sigma \rangle  \\ 
        &\quad + s_1 s_2 \left(a \cdot b\right) + s_1 s_3 \left(a \cdot c\right) + s_1 s_4 \left(a \cdot d\right) + s_2 s_3 \left(b\cdot c\right) + s_2 s_4 \left(b\cdot d\right) + s_3 s_4 \left(c \cdot d\right) \\ \nonumber
        &\quad + s_1 s_2 s_3 [ (a \cdot b) \langle c \cdot \sigma \rangle + (b \cdot c) \langle a \cdot \sigma \rangle - (a\cdot c) \langle b \cdot \sigma \rangle ] \\ \nonumber
        &\quad + s_1 s_2 s_4 [ (a \cdot b) \langle d \cdot \sigma \rangle + (b \cdot d) \langle a \cdot \sigma \rangle - (a\cdot d) \langle b \cdot \sigma \rangle ] \\ \nonumber
        &\quad + s_1 s_3 s_4 [ (a \cdot c) \langle d \cdot \sigma \rangle + (c \cdot d) \langle a \cdot \sigma \rangle - (a\cdot d) \langle c \cdot \sigma \rangle ] \\ \nonumber
        &\quad + s_2 s_3 s_4 [ (b\cdot c) \langle d \cdot \sigma \rangle + (c \cdot d) \langle b \cdot \sigma \rangle - (b\cdot d) \langle c \cdot \sigma \rangle ] \\ \nonumber
        &\quad + s_1 s_2 s_3 s_4 [ (a \cdot b) (c \cdot d) + (b \cdot c) (a \cdot d) - (a\cdot c) (b \cdot d) ] \} \nonumber
    \end{align}
\end{widetext}
where $\rho = \ket{\psi}\bra{\psi}$ is set to be a pure state and $\langle e_i \cdot \sigma\rangle \equiv \bra{\psi} e_i \cdot \sigma \ket{\psi}$. For the numerical simulation in Fig.2, the measurement angles are parametrised by a set of parameters $\{\zeta,\beta \}$ as
\begin{align}
    \theta_k &= \theta_{k,0} + \zeta_k \beta
    \\
    \gamma_m &= \gamma_{m,0} + \zeta_m \beta
\end{align}
where $\zeta \in \mathbb{R}$ is an arbitrary scaling number and $\beta \in [0,2\pi]$. Additionally, the initial angles $\{\theta_{i,0},\gamma_{i,0}\}$ are assigned fixed random values. Figure 2 shows the classical bound together with the generalised bounds of gLGI3 and gLGI4 as functions of $\beta$. It follows that no violation of the gLGIs occurs for
any of the measurement settings considered. This result is expected since the probability boundary is extended by allowing a negative joint probability, then there should be no violation. 
\\
\indent Even though the gLGIs are satisfied, it requires a reconsideration of the conditions underlying Macrorealism in the context of extended probability. The system now has a generalised MR (gMR) if the following assumptions are satisfied simultaneously:
\begin{itemize}
    \item \textit{generalised Macrorealism per se} (gMRps) --- The state solution possesses value at each time, for example a spin measurement, $s_i = \pm1$ at time $t_i$, independently whether it is being measured or not.
    \item \textit{Non-invasive extended probability} (NIEP) --- Extended probability at each time $t_i$ is independent of those at any other time $t_j$.
    \item \textit{Induction} --- A future measurement, whether measured or not, should have no effect on the current state. This assumption is always assumed to hold. This one remains unchanged.
\end{itemize}
A generalised MRps allows us to associate even non-settleable histories into consideration which is in contrast to the usual MRps where only settleable histories are considered. This  is a relaxation of the usual MRps. Moreover, since there is no need for any measurement to be actually taken, we reframe NIM by the independence in probability between times, namely NIEP, rather than a statement about measurement. Therefore, if the gLGIs are satisfied, then the quantum system can be said to possess generalised Macrorealism under the extended probability. Or one could say that the suitable probability in quantum system, for exhibiting classical perspectives, is arguably the extended probability.

\section{No violation in a generalised Bell-CHSH inequality with extended probability}
\indent We construct a generalised Bell-CHSH inequality by considering the definition of a modified spatial correlation between subsystems, $A$ and $B$. That is,
\begin{equation}
    \tilde{C}_{s_a s_b} = \sum_{s_a s_b} s_a s_b \; q(s_a,s_b)
\end{equation}
where the extended probability is
\begin{equation}
    q(s_a,s_b) = \text{ReTr} \left[ \left( P_{s_a} \otimes P_{s_b} \right) \rho \right].
\end{equation}
Similar to the gLGI case, this modified spatial correlation has the same value as the one with the classical probability. Following the same steps as in Sec.III, we obtain
\begin{align}
    \tilde{B}_4 = |\tilde{C}_{ab}& +  \tilde{C}_{a'b} + \tilde{C}_{ab'} -  \tilde{C}_{a'b'}|  \notag \\
    &\le \sum_{s_a s_{a'} s_b s_{b'}}|s_a s_b + s_{a'} s_b + s_a s_{b'} - s_{a'} s_{b'}| 
    \\ & \quad \quad \quad \quad \cdot |q(s_a,s_{a'},s_b,s_{b'})|. \notag
\end{align}
A generalised Bell-CHSH inequality hence reads 
\begin{equation}
    |\tilde{C}_{ab} + \tilde{C}_{ab'} + \tilde{C}_{a'b} - \tilde{C}_{a'b'}| \le 2 \mathscr{I}_4.
\end{equation}
\indent For a further analysis, we consider a simple EPRB model. The setting is a standard system which is an entangled singlet state of two electrons ($\ket{\psi}$) sending to Alice ($A$) and Bob ($B$). The system is being measured with the spin measurements ($P_s$) on an angle set $\{a,a',b,b'\}$ that associated with the parameters as in Fig.2. The extended probability takes the form
\begin{widetext}
    \begin{align}
        q(s_a,s_{a'},s_b,s_{b'}) &= \text{ReTr}\left( P_{s_a}^a P_{s_{a'}}^{a'} \otimes P_{s_b}^{b}P_{s_{b'}}^{b'} \ket{\psi}\bra{\psi} \right)
        \nonumber \\
        &= \dfrac{1}{16} \big\{ 1 + s_as_{a'} (a\cdot a') + s_as_b(a \cdot b) + s_as_{b'} (a \cdot b') \\ \nonumber
        &\quad + s_{a'}s_b (a' \cdot b) + s_{a'}s_{b'} (a'\cdot b') + s_bs_{b'} (b \cdot b') 
        \\ \nonumber
        &\quad + s_as_{a'}s_bs_{b'} \left[ (a\cdot a')(b\cdot b') + (a\cdot b')(a' \cdot b) - (a\cdot b)(a' \cdot b') \right] \big\}.
        \end{align}
\end{widetext}
This extended probability always satisfies the generalised Bell-CHSH inequality of Eq.(32) as shown in Fig.3, following the same numerical simulation as in Sec.III. Thus, this suggests a reconsideration of Local realism within the extended probability framework. We may introduce a generalised Local realism (gLR) which are
\begin{itemize}
    \item \textit{generalised Determinism} --- The state of a system contains definite properties independent of whether the measurement occurs.
    \item \textit{generalised Locality} --- The properties assigned to one subsystem is independent of local setting of the other spatially separated subsystem.
\end{itemize}
\indent Significantly, the derivation above does not require the introduction of a local hidden variable. Rather, it implies only the existence of the extended probability space. The notion of locality here is closer to what have been discussed in \cite{rothman2001hidden}, that is, a locality without a local hidden variable notion. The equivalence of the two pictures is possible via Fine's theorem \cite{fine1982joint,fine1982hidden,halliwell2019fine}. However, the satisfaction of Eq.(32) does not generally permits a local hidden variable notion as no generalised analogy of Fine's theorem has yet been established. Still, it is useful to examine the results within the local hidden variable framework as a possible representation for discussing the meaning of generalised determinism and generalised locality.
\\
\indent Locality is thus possibly supposed to be defined with a local hidden variable $\lambda$ notion,
\begin{align}
    q(A|a,b,\lambda) &= q(A|a,\lambda)
    \\
    q(B|a,b,\lambda) &= q(B|b,\lambda),
\end{align}
as well as the factorisability condition
\begin{equation}
    q(A,B|a,b,\lambda) = q(A|a,\lambda)q(B|b,\lambda).
\end{equation}
The generalised locality here is also nothing but an assumption, similar to the usual Locality in Sec.II.B. but formulated without requiring any measurement to occur. 
\\
\indent On the contrary, the generalised Determinism requires more discussions since it concerns the Counterfactual definiteness (CFD) \cite{hess2016counterfactual, hance2024counterfactual}. Therefore, we present three possible levels of acceptance of the CFD (or implicitly the generalised Determinism itself) ranging from a more classical to a quantum point of view
\begin{itemize}
    \item CFD amongst incompatible sets of CH is \textit{completely allowed} --- If we presume that the system completely behaves classically once a set of Eq.(32) is satisfied setting aside the single-framework rule, then the CFD is completely allowed. In this case, we can assign a (extended)probability value to performed or unperformed measurements at will. However, this presumption goes against the single-framework rule of the histories, which is the core principle of the CH framework.
    \item CFD amongst incompatible sets of CH is \textit{conditionally allowed} --- According to \cite{halliwell2017incompatible}, if one acquires a \textit{unifying probability} over non-commuting projection operators then an \textit{extended single-framework rule} is applicable. Certainly, the difference is the extended probability is applied entirely, not only for some range, in this case. This is a different class of history space that is not obligated by a single-framework rule, a class that we could observe multiple quantities of non-commuting observables simultaneously as in a classical realm.
    \item CFD amongst incompatible sets of CH is \textit{not allowed} --- If we respect the single-framework rule in the usual sense \cite{griffiths2020nonlocality}, a local hidden variable ($\lambda$) is perceived as a quantum variable in this perspective; that is, CFD is not applicable across incompatible frameworks. A single probability (or extended probability) space for incompatible observables $\lambda$ is impossible in a quantum system.
\end{itemize}
\indent The second point of view provides a possible way to reconcile classical descriptions and quantum descriptions. It implies that non-commuting quantities could be observed simultaneously once the generalised inequality in (32) is satisfied, which is normal in a classical perspective.
\\
\indent Counterfactual definiteness is guaranteed for a classical system whereas it is reluctant to be introduced in a quantum system. Since there is no direct influence between Alice and Bob based on the history framework, whether the underlying the local hidden variable ($\lambda$), or underlying mechanism, should be a classical or quantum variable remains as an open question in this point. The classicality here is expressed via an extended probability in the local hidden variable space, unlike in the gLGI case where the results still belong to the history space. 

\begin{figure}
    \centering
    \includegraphics[width=0.4\textwidth]{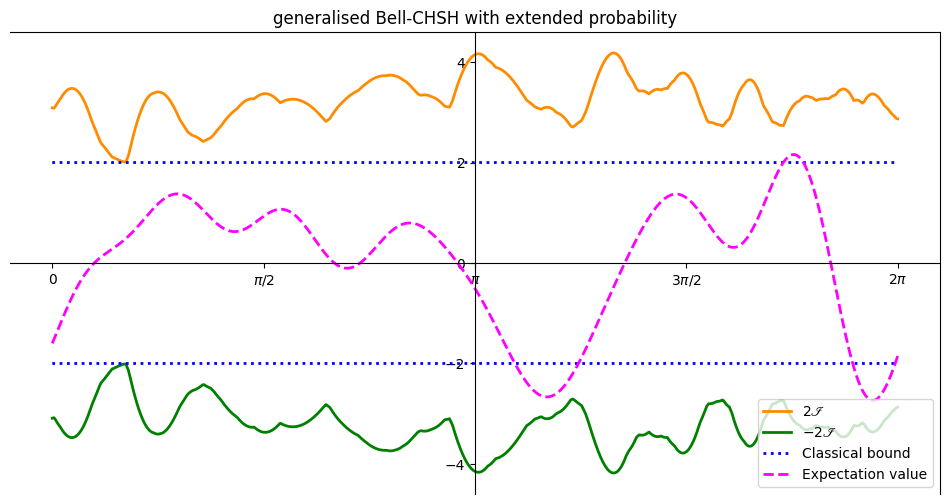}
    \caption{illustrates an extended bound for a representative measurement setting in CHSH showing no violation occurred.}
    \label{fig:B4}
\end{figure}

\section{Additional remark on extended unity}
In a standard Leggett-Garg inequality or Bell-CHSH inequality, the bound is captured by the fact that $\sum p=1$. This allows one to reduce one inequility to another. For example, LGI4$\rightarrow$LGI3 where the boundaries are perfectly matched. However, when it comes to an extended unity, the same reduction is no longer guaranteed. This is because $q$ is not confined to the interval $[0,1]$. Thus, It is possible that $\mathscr{I}_{4 \rightarrow 3} \ne \mathscr{I}_3$.
\\
\indent Let us take a projection operator as Eq.(25) and set two of the measurement angles in gLGI4 equal. The extended probability then reduces to
\begin{equation}
    q(s_1,s_2,s_3,s_4)_{s_3 = s_4} = \delta_{s_3,s_4} \text{ReTr}\left( P_{s_3}P_{s_2}P_{s_1} \rho \right).
\end{equation}
Obviously, the reduction of extended unity can be shown as
\begin{equation}
    \mathscr{I}_{4\rightarrow3} = \sum_{s_1s_2s_3} \left| q(s_1,s_2,s_3) \right| \sum_{s_4}\left| \delta_{s_3,s_4} \right| = \mathscr{I}_3.
\end{equation}
Nevertheless, the above is no longer true for the unsharp measurement that is characterised by an unsharp parameter $\eta$. Thus, a spin projection operator of each time has the form
\begin{equation}
    P_{s}^{e}(\eta) = \dfrac{1}{2} \left[ 1 + s \; \eta \;(e \cdot \sigma) \right]
\end{equation}
where $e$ is the measurement angle identical to the setting in Sec.III. and $0 \le \eta \le 1$. 
\\
\indent Since the projection operators now are no longer idempotent, i.e. $P_{s}^{e}(\eta)P_{s'}^{e}(\eta) \ne \delta_{s,s'} P_{s}^{e}(\eta)$, the multiplication of the identical projection operator generally takes the form
\begin{align}
    P_{s}^{e}(\eta)P_{s'}^{e}(\eta) = \left( P_{s}^{e} + \dfrac{1}{4}  \left( \eta^2 -1 \right) \right) \delta_{s,s'} 
    \\
    + \dfrac{1}{4} \left( 1-\eta^2\right)\delta_{s,-s'}. \notag
\end{align}
Therefore, the reduced extended unity of gLGI4 compared to extended unity of gLGI3 can be expressed as
\begin{align}
    |q(s_1,s_2,s_3)| &\le \sum_{s_4} |q(s_1,s_2,s_3,s_4)_{s_3 = s_4}| \notag
    \\
    \mathscr{I}_3 &\le \mathscr{I}_{4\rightarrow3}.
\end{align}
As shown in Fig.4, the numerical simulation reveals no equivalence of the inequality of (42) in terms of $\Delta \mathscr{I} = \mathscr{I}_{4\rightarrow3} - \mathscr{I}_3$ as a function of $\beta \in [0,2\pi]$. 
\\
\indent To summarise, the extended unity is sensitive to the unsharpness value while the standard unity is forgiven. This has never been the case for a classical probability since $\mathscr{I}_{\text{classical}} = 1$ always. Moreover, the reduction of the whole boundary is still preserved for the sharp projection operator setting, conversely, such a reduction is broken for the unsharp scenario. Note that the reduction of the extended probability alone is perfectly achievable.

\begin{figure}
    \centering
    \includegraphics[width=0.4\textwidth]{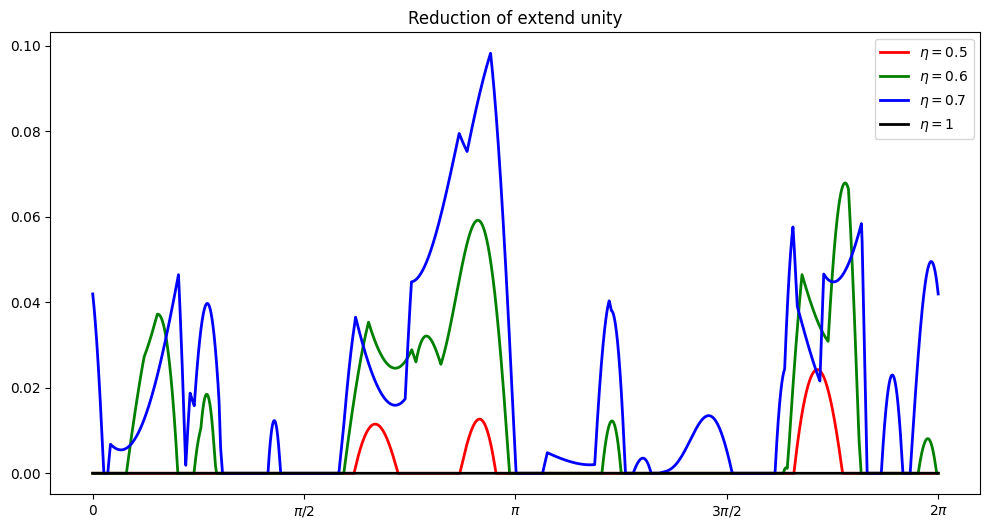}
    \caption{illustrates a reduction of extended unity at different value of unsharp parameter $\eta$.}
    \label{Extended unity}
\end{figure}

\section{Conclusion}
\indent The generalised Leggett-Garg inequality in (23) and generalised Bell-CHSH inequality in (32) are presented respectively by replacing a classical probability notion with extended probability notion. In other words, the extension of the bounds are captured by an extended unity in Eq.(24). Importantly, the generalised inequalities exhibit no violation for arbitrary measurement settings we consider. Since the extended probability is not associated with an measurement process, an observation or analysis that related to measurement needs to be relaxed or taken with care. Despite such a caveat, the extended probability seems to be compatible in principle to a quantum system while showing classicalities more than a classical probability.
\\
\indent The extended probability is adapted into the consistent history framework pioneered by  J.B. Hartle \cite{hartle2008quantum, hartle2016decoherent}. In this picture, the quantum system is a closed system that can be described within two layers of reality. A fundamental layer associated with a the non-settleable history which is expressed with extended probability, whilst an observable layer is described by a classical probability. The two layers are connected via a decoherence functional, i.e. the interference between history chains. Thus, if one requires an observable history then a suitable decoherence condition is applied \cite{diosi2004anomalies, halliwell2009partial}. The whole generalisation of our model relies on the fact that a quantum system is describable with such a fundamental layer using an extended probability.
\\
\indent An absence of violation of the generalised Leggett-Garg inequality implies that a quantum system follows a generalised Macrorealism. This is a generalised notion of MR such that no measurement is required. Then, generalised Macrorealism per se is acquired when the system possesses an definite property at any time. This is in agreement with the picture of the consistent history framework where non-settleable histories are not included. The main difference from the traditional MR is the non-invasive extended probability. Although there is no an act of measurement, the quantum system can still be time-independent in an extended probability notion. NIEP characterises this property, which the extended probability always satisfies it.
\\
\indent Similarly, for the generalised Bell-CHSH inequality, no violation implies that the quantum system possesses a generalised Local realism. That is, the generalised Locality is formulated as an assumption within the consistent history framework. It concerns the probabilistic independence of spatially separated subsystems even when no measurement is performed. On the contrary, the generalised Determinism is more subtle when exploiting within the history framework.  It concerns the Counterfactual definiteness (CFD) \cite{hess2005bell,hance2024counterfactual}. We have discussed three levels to which such a classicality can be associated: CFD is completely allowed, CFD is conditionally allowed, CFD is not allowed. Our discussion is based on a local hidden variable which is deduced to be possible once the generalised Bell-CHSH inequality is satisfied. 
\\
\indent In conclusion, the quantum system can admit a generalised notion of classicality that is characterised by a generalised Leggett-Garg inequality and a generalised Bell-CHSH inequality. In this regard, the extended probability seems suitable for the quantum system to recover classical perspectives. Nevertheless, how valid is the generalisation of this kind? Honestly, it should be asked ``How valid is the extended probability?" instead. This question has a long history of pursuit. It is moderately acceptable that physicists take it as a calculation tool that facilitates a demonstration of some physical features. Our model is one of the examples that utilises this fact. If the quantum system with extended probability has a generalised Fine's theorem, all of these struggles will be presumably mitigated. This is one of the possible further development on this project which will be explored elsewhere.

\section*{Acknowledgement}
This work was supported by the National Science, Research and Innovation Fund (NSRF) via the Program Management Unit for Human Resources and Institutional Development, Research and Innovation (grant No.B39G690007). S.K. and P.S. acknowledge support from the NSRF under the aforementioned grant.

\bibliography{references}
\end{document}